\documentclass[aps,prev,twocolumn,preprintnumbers,floatfix,nofootinbib]{revtex4-1}
\usepackage{graphicx}
\usepackage{bm}
\usepackage{times}
\usepackage[hypertex]{hyperref}
\usepackage{slashed}
\def\br{\begin{eqnarray}}
\def\er{\end{eqnarray}}
\def\be{\begin{equation}}
\def\ee{\end{equation}}

\def\({\left(}
\def\){\right)}

\begin{document}

\title{On the connection of Gamma-rays, Dark Matter and Higgs searches at LHC}

\author{J. D. R. Alvares$^{b}$}\email{jose@udea.edu.co}
\author{C. A. de S. Pires$^{a}$}\email{cpires@fisica.ufpb.br}
\author{Farinaldo S. Queiroz$^{a,c}$}\email{fqueiroz@fnal.gov}   
\author{D. Restrepo$^{b}$}\email{restrepo@udea.edu.co}
\author{P. S. Rodrigues da Silva$^{a}$}\email{psilva@fisica.ufpb.br}

\affiliation{$^{a}$Departamento de F\'{\i}sica, Universidade Federal da
Para\'{\i}ba, 
Caixa Postal 5008, 58051-970, Jo\~ao Pessoa - PB, Brazil.
\\
$^{b}$ Instituto de Fisica, Universidad de Antioquia, A.A. 1226, Medellin, Colombia.
\\
$^c$Center for Particle Astrophysics, Fermi National Accelerator Laboratory, Batavia, IL 60510, USA
}

\date{\today}
\vspace{1cm}
\begin{abstract}

Motivated by the upcoming Higgs analyzes we investigate the importance of the complementarity of the Higgs boson chase on the low mass WIMP search in direct detection experiments and the gamma-ray emission from the Galactic Center measured by the Fermi-LAT telescope in the context of the $SU(3)_c\otimes SU(3)_L\otimes U(1)_N$.  We obtain the relic abundance, thermal cross section, the WIMP-nucleon cross section in the low mass regime and network them with the branching ratios of the Higgs boson in the model. We conclude that the Higgs boson search has a profound connection to the dark matter problem in our model, in particular for the case that ($M_{WIMP} < 60$ GeV ) the BR($H \rightarrow 2$ WIMPs ) $\gtrsim 90\%$. This scenario could explain this plateau of any mild excess regarding the Higgs search as well as explain the gamma-ray emission from the galactic center through the $b\bar{b}$ channel with a WIMP in the mass range of $25-45$~GeV, while still being consistent with the current limits from XENON100 and CDMSII. However, if the recent modest excesses measured at LHC and TEVATRON are confirmed and consistent with a standard model Higgs boson this would imply that $ M_{WIMP} > 60 $~GeV, consequently ruling out any attempt to explain the Fermi-LAT observations.

\end{abstract}

\maketitle
\section{Introduction}

The nature of the Dark Matter (DM) is one of the biggest mysteries of the universe and 
lies on the interface of particle physics, astrophysics and cosmology. It is common 
sense that, in order to determine its nature, a complementary search in direct and indirect 
detection plus collider experiments is necessary. The long standing DAMA/LIBRA experiment 
reports with significance of $8.9\sigma$ the detection of an annual modulation with a phase 
and period consistent with elastically scattering dark matter, a WIMP (Weakly Interacting Massive Particle)~\cite{DAMA} with a mass of $\sim 7$~GeV, with a WIMP-nucleon cross section of 
 $\sim 10^{-41}cm^2$~\cite{Chris1}.  The CRESST and CoGeNT experiments observed recently
  some excess events consistent with WIMP scattering off a nuclei with similar spectrum of
   events~\cite{CRESST,CoGeNT}. The CoGeNT collaboration has reported an annual modulation with an 
   amplitude higher than DAMA but also consistent with the WIMP hypothesis. Those observations seem to point to an imminent WIMP discovery in the near future. However the DAMA modulation
is arguable since there are analyses which claim that this modulation could be due to 
    cosmic ray muons~\cite{KfirBlum}, while others disagree with the muon hypothesis by 
    declaring that the phases are $\sim 3\sigma$ off~\cite{Bernabei} and the scattering 
    rates are different. With respect to the CoGeNT excesses, the uncertaints in the rise 
    time cut may result in a sizable contamination of surface events in CoGeNT data and 
    therefore a fraction of these excess events is expected to be residual surface events~\cite{Chris1}. Lately a large background contamination is expected in the CRESST detector by $^{206}Pb$ 
decays and $\alpha$ particles \cite{CRESSTback}. Furthermore these signals appear to be 
in conflict with the other experiments such as CDMSII and XENON10. In particular, the CDMS collaboration has searched for this modulation in their last run but no annual modulation
was found~\cite{CDMSmodullation}. Conversely, the XENON detector which measures ionization 
and scintillation and has the strongest constraints in this mass range, suffers non-negligible 
uncertainties in the scintillation efficiency at low recoil energies as discussed in \cite{Juancollar}, 
making it difficult to interpret their resulting limits at low energy range. Likewise, it is important 
to mention that at such low energies the backgrounds observed by CDMS are not well understood, 
somewhat limiting their ability to probe the region implied by CoGeNT and DAMA-LIBRA~\cite{JuanCDMS}.
Be that as it may, one could evoke non-Maxwellian distributions and/or tidal streams to alleviate 
the tension among those experiments~\cite{Chris2}. In summary the claim wether these signals reported 
recently are due to dark matter or not is still debatable. Besides this search for dark matter 
at low energy recoils, an important and enlightening mono-photon and mono-jet search has been 
performed at the LHC and the strongest limits in the $\sim 1$~GeV mass window have been 
put for the case of light mediators~\cite{monojet}. Moreover an interesting and promising search for dark matter
is ongoing in a variety of ground based and space based telescopes~\cite{IndirectTelescopes}. 
In particular the Fermi Gamma Ray Telescope have been collecting data for almost four years from 
the region surrounding the center of the Milky Way which is both astrophysically rich and complex, 
and is predicted to contain very high densities of dark matter. By analyzing the morphology and 
spectrum of the gamma ray emission from this region, several groups~\cite{GCDM1}-\cite{GCDM2} have 
found an evidence of a spatially extended component which peaks at energies between $300$~MeV and $10$~GeV which can be either explained by the annihilations of dark matter particles in the inner galaxy, 
or through collisions of high energy protons (that are accelerated by the Milky Way's supermassive 
black hole) with gas. If interpreted as dark matter annihilation products, the emission spectrum 
favors dark matter particles with a mass in the range of $(25-45)$~GeV which annihilates mainly to 
$b\bar{b}$ (see Fig 6 of Ref.~\cite{GCDM2}).  Besides all these direct and indirect detection signals we 
have the ongoing Higgs analyses which soon will shed some light in the new physics models. Here we 
explore the complementarity of these three approaches in the context of the $SU(3)_C\otimes SU(3)_L\otimes U(1)_N$, $331_{LHN}$ for short~\cite{331models1}-\cite{331models4}. Besides featuring many of the standard model (SM) virtues it addresses fundamental questions such as the dark matter signals~\cite{CDM331} as well as many theoretical questions, such as number of families~\cite{331models1} and neutrinos masses, among others~\cite{331model3}. We will not dwell on the nice features of the model, but we recommend those works aforementioned for those who are interested in more detailed descriptions. Despite the cold dark matter problem has been already investigated previously in the context of 331 models in Ref.~\cite{CDM331}, here we will also investigate indirect detection signals, derive the branching ratios of the Higgs boson into WIMPs, $b \bar{b}$, $\gamma \gamma$ and $\tau \bar{\tau}$ and examine the role of Higgs boson in direct and indirect detection searches by analyzing the impact of the ongoing Higgs boson chase on the parameter space of the model, such as to explain the gamma-ray emission from the galatic center, while obeying the current bounds coming from direct detection experiments such as XENON100 and CDMSII.

We shortly describe the model in section~\ref{sec1} by introducing its main ingredients. In section~\ref{sec3} we discuss in more details the direct and indirect evidences for dark matter as well as our reasonings. Further in section~\ref{sec4} we discuss the impact of the Higgs boson search on our results and network them with the Dark Matter problem. Lastly, we present our conclusions.

\section{The 3-3-1LHN Model}
\label{sec1}

Our framework is the  $331_{LHN}$ model~\cite{331models1}, which is a direct extension of the electroweak sector of the SM. 
In order to allow the reader to follow our reasonings we will briefly discuss the content of the model hereafter.

\subsection{Fermionic content}

Likewise the SM, the leptonic sector is placed with left-handed fields appearing in triplets, $f_{aL}=(\nu^a_L,l^a_L,N^a_L)^T$ transforming as $(1, 3,-1/3)$ and right-handed ones in singlets, $e_{a R}$  as $(1,1,-1)$ and $N^a_R$ as $(1,1,0)$ , where $a=1,2,3$ correspond to the three families. In the hadronic sector, the first two families are placed as anti-triplets $Q_{i L} = (d_{i L}, -u_{i L}, d^{\prime}_{i L} )^T$ as $(3,\bar{3},0)$ , with $i=1,2$,  while the third one is arranged as triplet, $Q_{3 L} = (u_{3 L}, d_{3 L}, u^{\prime}_{3 L} )^T$ as $(3,3,1/3)$ and, the right-handed quarks are singlets with hypercharges equal to their electric charges similarly to the SM. The first two and the third family of left-handed quarks are in different representations due to an anomaly cancellation requirement adequately described in previous works~\cite{331models1}. The primed fermions are the exotic ones, singlets under the SM gauge group. Similarly to the SM, all fermions acquire Dirac mass terms through a spontaneous symmetry breaking mechanism in the Higgs sector presented hereunder.

\subsection{Scalar content}

It was noticed that by introducing a global symmetry $U(1)_G$ where,
\begin{equation}
\mathbf{G}({\bar N}_{L/R},\,{\bar u}_{3L/R}^\prime,\,d_{iL/R}^\prime,\,V_\mu^-,\,U_\mu^{0},\,\chi^{0},\,\chi^{-},\,\eta^{\prime 0 *},\,\rho^{\prime -})=+1\,,
\label{LNA}
\end{equation}
that we could avoid undesirable mixings in the gauge and scalar sector and, in addition, obtain the lightest particle charged under this symmetry to be stable. All this procedure was already discussed in details in Ref.~\cite{CDM331}, hence we skip it here. In summary, we introduce three scalar triplets, namely,
\begin{eqnarray}
\label{conteudoescalar}
\chi & = & (\chi^0, \chi^-, \chi^{\prime 0})^T,\nonumber \\
\rho & = & (\rho^+, \rho^0, \rho^{\prime +})^T,\nonumber \\
\eta & = & (\eta^0, \eta^-, \eta^{\prime 0})^T,
\end{eqnarray}along with the following Yukawa Lagrangian,

\begin{eqnarray}
&-&{\cal L}^Y =f_{ij} \bar Q_{iL}\chi^* d^{\prime}_{jR} +f_{33} \bar Q_{3L}\chi u^{\prime}_{3R} + g_{ia}\bar Q_{iL}\eta^* d_{aR} \nonumber \\
&&+h_{3a} \bar Q_{3L}\eta u_{aR} +g_{3a}\bar Q_{3L}\rho d_{aR}+h_{ia}\bar Q_{iL}\rho^* u_{aR} \nonumber \\
&&+ G_{ab}\bar f_{aL} \rho e_{bR}+g^{\prime}_{ab}\bar{f}_{aL}\chi N_{bR}+ \mbox{H.c}, 
\label{yukawa}
\end{eqnarray}
where the triplets $\eta$ and $\chi$ both transforming as (1,3,-1/3) and $\rho$ as (1,3,2/3) and in Eq.\ref{yukawa} we are using the family indexes $i=1,2$ and $a=1,2,3$.

For our analyses, the most important feature in this model is that the $U(1)_G$ symmetry implies that the charged particles under this symmetry are produced in pairs,  like R-parity in supersymmetric theories\cite{SUSYR}, and hence the lightest typical 331LHN model's particle will be stable. In this way the model can provide two (non-simultaneous) WIMP candidates, $N_1$ and $\Phi$, where $N_1$ is a heavy neutrino and $\Phi$ is a complex neutral scalar, which arises from the combination of the $\chi^0$ and $\eta^{0 \prime}$ scalar fields after the diagonalization procedure~\cite{CDM331}.
Here we will explore only the case where the scalar $\Phi$ is the dark matter candidate. The reason relies on the fact that the other possible WIMP, $N_1$, has an excluded WIMP-nucleon cross section in the low mass regime, which is exactly the region of mass we are interested in this work. All this being said, we will use the terminology WIMP to refer to our scalar $\Phi$ from now on. In the next section we will investigate the status of our model with respect to the direct and indirect detection searches of dark matter.

\section{Direct and Indirect Detection} 
\label{sec3}

Among the CDM candidates, the WIMPs are the most promising ones for providing a thermal cross section roughly at the electroweak scale, naturally leading to the appropriate relic density, as well as because the current and next generation of direct and indirect detection experiments are sensitive to the parameter space where most of the theoretical models rely on. Before examining the status of our model concerning those searches, we will scan the parameter space of the model and check if our WIMP can account for the total observed dark matter abundance by computing,
\begin{equation} 
\Omega h^{2}=2.742\times 10^{8}\frac{M_{WIMP}}{\mbox{GeV}} Y(T_{0}),
\end{equation}
where $Y(T_{0})$ is the number density over entropy evaluated today.

\begin{figure}[!htb]
\centering
\includegraphics[width=0.8\columnwidth]{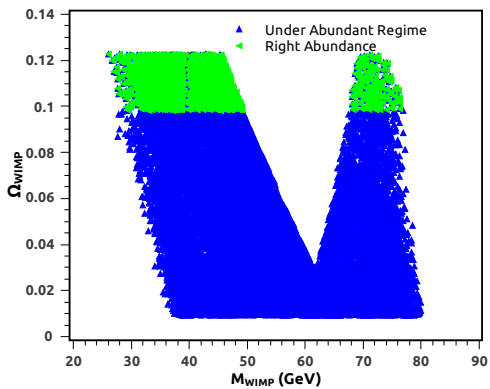}
\caption{Abundance of the WIMP ($\Phi$) as function of its mass. Green (blue) scatter refers to points where the WIMP provides the correct abundance (is under abundant). Correct abundance means $ 0.098\leq \Omega h^2 \leq 0.122$ while the under abundant regime is for $0.01 \leq \Omega h^2 \leq 0.098$. }
\label{AbundanceMfi}
\end{figure}

\begin{figure}[!htb]
\centering
\includegraphics[width=0.8\columnwidth]{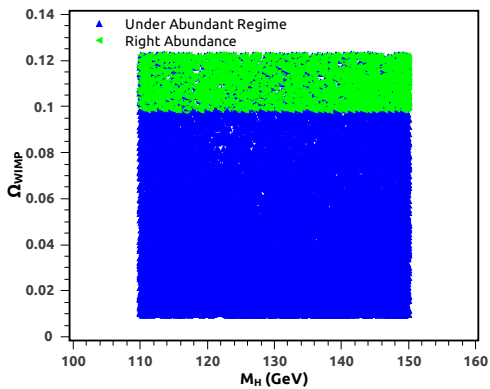}
\caption{ Abundance of the WIMP  ($\Phi$) as function of the Higgs boson mass ($M_{H}$). Green (blue) scatter refers to points where the WIMP provides the correct abundance (is under abundant). Correct abundance means $ 0.098\leq \Omega h^2 \leq 0.122$ while the under abundant regime is for $0.01 \leq \Omega h^2 \leq 0.098$. }
\label{AbundanceMH}
\end{figure}

We used the Micromegas package  where we implemented the model in, to take into account all the processes which contribute to the relic abundance of our WIMP, $\Phi$, automatically~\cite{micromegas}. These processes were explicitly exhibited in Ref.~\cite{CDM331}. In FIG.~\ref{AbundanceMfi} we show the abundance of the WIMP as function of its mass and in FIG.~\ref{AbundanceMH} as a function of the mass of Higgs boson in the low mass WIMP regime, $M_{WIMP} < 60$~GeV.

From FIG.~\ref{AbundanceMfi} we can clearly see that our model has a largish region where our WIMP, $\Phi$, provides  the right abundance according to WMAP7~\cite{WMAP7}. From FIG.~\ref{AbundanceMH} we cannot see any bias towards a light Higgs boson ($M_{H} < 130$~GeV) and therefore by looking at the abundance dependence on the Higgs boson mass only, we cannot extract any information relevant to the Higgs boson search differently from some doublet Higgs models~\cite{refmissing}. 
Now that we have confirmed that our  dark matter candidate can reproduce the right abundance, we will investigate if our model is also consistent with current limits from direct detection experiments. 

Since we have a flux of WIMPS surrounding us we expect to observe these WIMPs by detecting WIMP-nucleon scatterings in underground detectors. The measured quantities vary according to the detector technology, be that as it may, after making some assumptions concerning the dark matter distribution, all of them convert their results into the simple cross section (in the spin-independent case), 
\begin{equation}
\sigma_0 = \frac{4 \mu_r^2}{\pi}\left( Z f_p + (A-Z) f_n \right)^2\,,
\end{equation} 
where $Z$ is the atomic number, $A$ is the atomic mass and $f_{p}$ and $f_{n}$ are effective couplings with protons and neutrons, respectively, and depends on the particle physics input of a given model. It is important to emphasize that these couplings are obtained numerically in our model by the MicrOMEGAs package by following the recipe described in Ref.~\cite{micromegas}. 

As aforementioned, the direct detection signals observed by CoGeNT, CRESST and DAMA may not be due to WIMP scatterings, so under the null hypothesis we are only concerned whether our WIMP candidate has a WIMP-nucleon cross section below the current bounds. 
\begin{figure}[!htb]
\centering
\includegraphics[width=0.8\columnwidth]{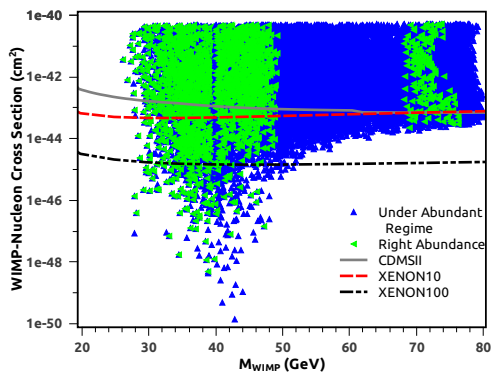}
\caption{ WIMP-nucleon cross section as function of the WIMP mass. Green (blue) scatter refer to points where the WIMP provides the correct abundance (is under abundant). Correct abundance means $ 0.098\leq \Omega h^2 \leq 0.122$ while the under abundant regime is for $0.01 \leq \Omega h^2 \leq 0.098$. }
\label{CSMfi}
\end{figure}

\begin{figure}[!htb]
\centering
\includegraphics[width=0.8\columnwidth]{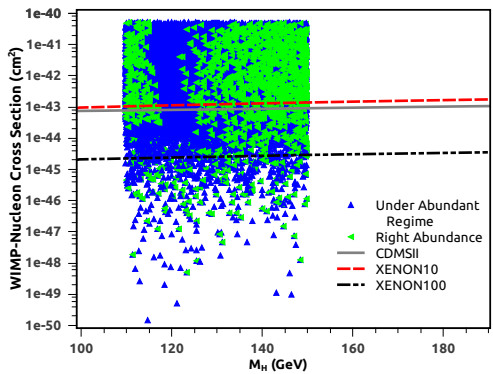}
\caption{ WIMP-nucleon cross section as function of the Higgs mass ($M_{H}$).Green (blue) scatter refer to points where the WIMP provides the correct abundance (is under abundant). Correct abundance means $ 0.098\leq \Omega h^2 \leq 0.122$ while the under abundant regime is for $0.01 \leq \Omega h^2 \leq 0.098$. }
\label{CSMH}
\end{figure}

In FIG.\ref{CSMfi}-\ref{CSMH} we have plotted the WIMP-nucleon cross section as function of the WIMP and Higgs boson masses, respectively. It may be noticed that the majority of the points are excluded by the XENON100 experiment, however we still have a worthwhile region which is completely consistent with the current limits. It is important to emphasize though, that the ongoing SuperCDMS at SNOWLAB, which will have a larger exposure by increasing the mass of the detector plus a better handle in discriminating surface events for implementing the new izip Germanium detectors, and the XENON1T for basically having a larger exposure, will be crucial to test our model in the near future, since they expect to improve their limits by roughly an order of magnitude.

In order to pin down the properties of the dark matter particle and find out the nature of the dark matter, we should also search for dark matter annihilations in our neighborhood.  It has been thought that if a sizable fraction of dark matter particles can annihilate into a pair of SM particles, the Galactic Center would be one of the best place to search for, since it would be the brightest in gamma-ray emission and for providing better statistics. 

The flux of gamma-rays coming from DM annihilation is given by,
\begin{equation}
\phi_{\gamma}(E_\gamma, \psi) = \frac{dN_{\gamma}}{dE_{\gamma}} \frac{ \left\langle  \sigma v \right\rangle }{8 \pi M^{2}_{WIMP} }\int_{los} \rho^2(r) dl,
\label{flux}
\end{equation}
where $\left\langle  \sigma v \right\rangle$ is the dark matter annihilation cross section times the relative velocity of the incoming WIMPs averaged over the velocity distribution and $\psi$ is the angle observed relative to the direction of the Galactic Center. The dark matter density as a function of distance to the Galactic Center (GC) is given by $\rho(r)$, and the integral is performed over the line-of-sight. $\frac{dN}{dE_\gamma}$ is the gamma-ray spectrum generated per annihilation.

In the right hand side of Eq.~(\ref{flux}) we have two different informations. The integration is the astrophysical input while the other terms refer to the particle physics information, and therefore model dependent. It is noteworthy to point out that the particle physics information is essentially the mass of the dark matter particle and the value of the cross section into the final states we are interested in. Hence once you measure the flux and assume a dark matter distribution and decay mode, there are two free parameters left, which are the mass of the dark matter particle and  the thermal cross section. Based on this information, we will investigate the possibility of explaining the Gamma-ray emission from the GC in our model.

\begin{figure}[!htb]
\centering
\includegraphics[width=0.8\columnwidth]{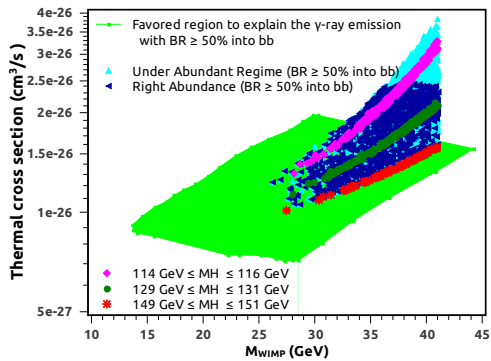}
\caption{Annihilation cross section as function of $M_{WIMP}$ for $BR(b\bar{b}) \geq 50\%$. Green region represents the favored region by the gamma-ray emission detected by Fermi-LAT in the GC. Dark (light) blue  points refer to the case where the WIMP provides the correct abundance (is under abundant). All (dark+light) blue points are for $110\ \mbox{GeV} \leq M_H \leq 150\ \mbox{GeV}$. Correct abundance means $ 0.098\leq \Omega h^2 \leq 0.122$ while the under abundant regime is for ($0.01 \leq \Omega h^2 \leq 0.098$). See the text for details.}
\label{ThermalCSMfi}
\end{figure}

By analyzing the Fermi-LAT data from August 4th, 2008 and August 3rd, 2011 and using the ULTRACLEAN class of events (events with less contamination of cosmic rays), a group has concluded that after subtracting the point source emission and the cosmic rays background, a residual emission from the inner 5 degrees surrounding the Galatic Center was present~\cite{GCDM2}, which is in good agreement with previous works~\cite{GCDM1}.
A number of proposals have been put forth to explain this gamma-ray emission. Since the morphology of the gamma-ray emission is not entirely point like, the black hole hypothesis might be ruled out, and because 44 out of 46 of the resolved millisecond pulsars by Fermi-LAT have a spectrum index larger than one, is somewhat unlikely to explain this gamma-ray emission through millisecond pulsars,because in order to explain this observed gamma-ray flux a large population of pulsars with a hard spectral index (~ $1.0$) would be required.
As we aforementioned this is not supported by the current data. 
By using a Navarro-Frank-White (NFW) \cite{micromegas} profile with an inner slope of $\gamma=1.3$,  which is suggested by cosmological simulations of Milky Way sized halos~\cite{MilkyWay},  it was concluded that $70\%$ up to $100\%$ of this gamma-ray emission is due to dark matter annihilations. In order to explain this gamma-ray emission a dark matter particle should annihilate mainly to $b\bar{b}$ and have a mass in the $\sim (15-45)$~GeV range as well as a thermally averaged annihilation cross section of ~$10^{-26} cm^3/s$ represented in the green region in FIG.~\ref{ThermalCSMfi}. In FIG.~\ref{ThermalCSMfi} we show the thermal cross section of our WIMP candidate as function of its mass for the case that it annihilates dominantly to $b\bar{b}$ ($> 50\%$)~\footnote{By modifying the main code in MicrOMEGAs we where able to store only the points where the annihilation to $b\bar{b}$ was the dominant one.}.  
From FIG.~\ref{ThermalCSMfi} we notice that our WIMP might be the origin of the gamma-ray emission from the GC and, at the same time, explain the dark matter abundance indicated by WMAP7 within the $(25\ \mbox{GeV} \leq M_{WIMP} \leq 40\ \mbox{GeV})$ mass range as well as that we cannot conclude anything concerning the dependence on the Higgs mass once the three regions: $114\ \mbox{GeV} \leq M_H \leq 116\ \mbox{GeV}$, $129\ \mbox{GeV} \leq M_H \leq 131\ \mbox{GeV}$ and $149\ \mbox{GeV} \leq M_H \leq 151\ \mbox{GeV}$ provide an equality good fit to the favored region to explain the gamma-ray emission from the GC. 

So far we have proved that our model is consistent with the current measurements regarding direct detection on the WIMP-nucleon cross section in the low mass regime as well as reproduce the right abundance indicated by WMAP7.  Furthermore we have shown that our model has a scalar WIMP, namely $\Phi$, which can explain at least the majority of gamma-ray emission coming from the GC. Now we will explore the complementarity of the ongoing Higgs boson search with the results we have discussed previously.

\section{Higgs connection} 
\label{sec4}

\begin{figure}[!htb]
\centering
\includegraphics[width=0.8\columnwidth]{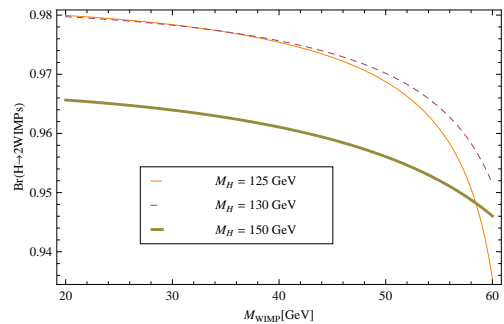}
\caption{Branching ratio of the Higgs boson into a pair of WIMPs. The solid thin orange line is for $M_H =125$~ GeV, the dashed purple line is for $M_H =130$~ GeV, the thick brown one is for $M_H =150$~ GeV. For the case that $M_{H}=125$~GeV and $M_{WIMP}=20$~GeV, $BR(H \rightarrow \gamma \gamma) \simeq 2.9 \times 10^{-5} $, $BR(H \rightarrow b \bar{b}) \simeq 1.4 \times 10^{-2} $, and $BR(H \rightarrow \tau \bar{\tau}) \simeq 7.9 \times 10^{-4} $.  We did not show these components in the plot for a matter of visualization.}
\label{GammaHfifi}
\end{figure}

Since we are at the LHC era the complementarity has become a promising and achievable way to 
shed some light in new physics models and, most importantly, disentangle 
them from other models. Therefore, we will network the ongoing 
Higgs measurements with the gamma-ray emission from the Galactic center 
and the bound coming from leading direct detection experiments, CDMS 
and XENON100. The Higgs boson is the only particle predicted by 
the SM of particle physics that has not yet 
been experimentally observed. Its observation would be a major 
step forward in our understanding of how particles acquire mass. 
Conversely, not finding the SM Higgs boson at the LHC and TEVATRON 
would be very intriguing  and would lead to
a greater focus on alternative theories that extend beyond 
the SM, with associated Higgs-like particles such 
as the 331 class of models.
While at TEVATRON the Higgs associated production with $b\bar{b}$ 
in the final state is the most important channel to look for the Higgs, 
at the LHC is the Higgs production via gluon fusion with $\gamma \gamma$ 
in the final state. The latter because of its great invariant mass resolution 
and efficient background rejection. Another channel that has been pursued is the 
$\tau \bar{\tau}$, for providing a good signal to noise ratio~\cite{CMS}. 
Since previous and current collider experiments have not observed any
mild excess consistent with a SM model Higgs boson, LEPII, TEVATRON and LHC 
have excluded the mass range $M_H < 115$~GeV and $M_H = 127-600$~GeV for a SM
Higgs boson. This plateau can be interestingly explained if the Higgs boson decays 
dominantly to a pair of DM particles, particularly WIMPs~\cite{HiggsDM}. This is exactly what happens in our
model for the case that $M_{WIMP} < 60$~GeV. In FIG.~\ref{GammaHfifi} we exhibit the
branching ratio for the Higgs boson into a pair of WIMPs as a function of the WIMP mass
for different Higgs masses (see analytical expression in Appendix \ref{sec6}). We can confirm from FIG.~\ref{GammaHfifi} that the Higgs boson
decays with a branching ratio larger than $90\%$ into WIMPs. For the case that $M_{WIMP} = 20$~GeV and $M_H=125$~GeV we obtained $BR(H \rightarrow \gamma \gamma) \simeq 2.9 \times 10^{-5} $ and $BR(H \rightarrow b \bar{b}) \simeq 1.4 \times 10^{-2} $, and $BR(H \rightarrow \tau \bar{\tau}) \simeq 7.9 \times 10^{-4} $ .

However this scenario would require an even more struggling search at the LHC mainly because the Higgs decays dominantly to missing energy in the final state and the $BR(H \rightarrow \gamma \gamma,b \bar{b},\tau \bar{\tau})$ would be suppressed~\footnote{We have added all possible decay channels for the Higgs boson including radiative decays such as $Z\gamma$ in order to derive precise values for the branching ratios}. On the other hand this scenario links strongly one of the most important problems in the modern science, which is the nature of the dark matter, with the Higgs boson paradigm which soon will unveil where the physics beyond the SM lies on.

It is worthy mentioning that this scenario is completely consistent with the current measurements. One may wonder if the well measured invisible width of the Z boson could constrain our model. Nevertheless, the Z boson cannot decay to WIMPs in our model simply because all the decay modes which involve a WIMP in the final state has at least one 331 particle that is much heavier than the Z boson.

\begin{figure}[!htb]
\centering
\includegraphics[width=0.8\columnwidth]{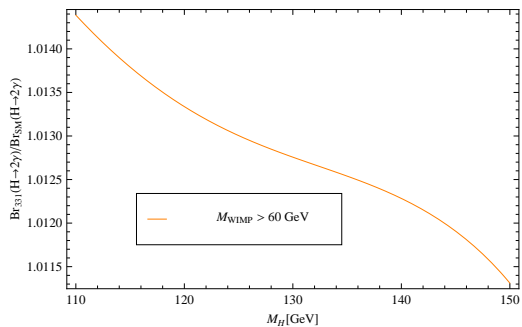}
\caption{Ratio of the branching ratios $H \rightarrow \gamma \gamma$ in the 331 model and in the SM for the case that $M_{WIMP} > 60$~GeV. }
\label{BRHAA}
\end{figure}

\begin{figure}[!htb]
\centering
\includegraphics[width=0.8\columnwidth]{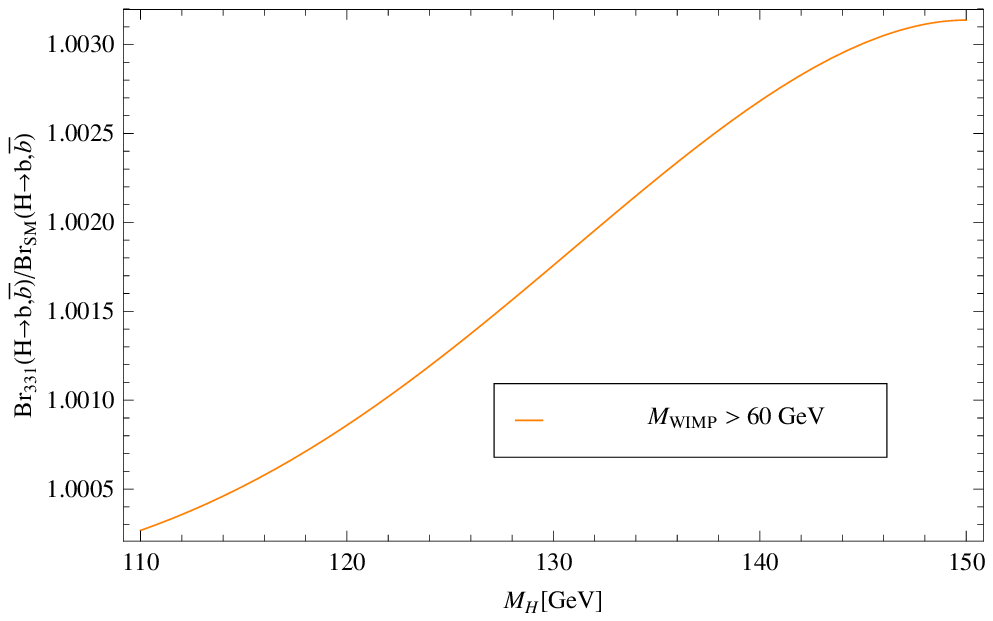}
\caption{Ratio of the branching ratios $H \rightarrow b \bar{b}$ in the 331 model and in the SM for the case that $M_{WIMP} > 60$~GeV.  }
\label{GammaHbb}
\end{figure}

\begin{figure}[!htb]
\centering
\includegraphics[width=0.8\columnwidth]{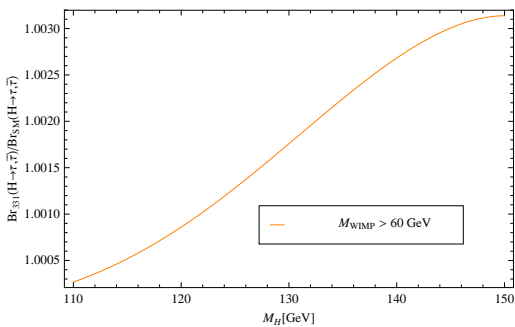}
\caption{Ratio of the branching ratios $H \rightarrow \tau \bar{\tau}$ in the 331 model and in the SM for the case that $M_{WIMP} > 60$~GeV.  }
\label{BRHtautau}
\end{figure}

Conversely, if the recent modest excesses measured at LHC and TEVATRON~\cite{Higgsexcess} are confirmed in the near future and consistent with a SM Higgs boson this completely rules out the DM hypothesis as an explanation to the 
gamma-ray emission from GC in our model, once this implies that $M_{WIMP} > 60 GeV$. This is so because only in the case that $M_{WIMP} > 60 GeV$ our model recovers the SM predictions regarding the Higgs branching ratios into $\gamma \gamma$, $b \bar{b}$ and $\tau \bar{\tau}$ as we can check in the FIGs.~\ref{BRHAA} to \ref{BRHtautau}, since from these plots we can clearly confirm that $BR_{331}/BR_{SM} \simeq 1$. We have focused in these channels for simplicity and for being the most promising channels to search for a light Higgs boson at LHC and TEVATRON. 

In summary the ongoing Higgs search has a complementary and crucial role in identifying the nature 
of the dark matter in our model since the properties of the Higgs boson are tightly related to the mass of the DM particle in the 331LHN model.

\section{Conclusions}
\label{sec5}

We have probed the low mass WIMP window, $M_{WIMP} < 60$~GeV, in the 331LHN model 
and checked if it can reproduce the right abundance of dark matter 
inferred by WMAP7 satellite in FIGs.~\ref{AbundanceMfi}~-~\ref{AbundanceMH}. Subsequently, we 
have shown that a sizable region of the parameter space is constrained by the current bounds 
derived by the XENON100 and CDMSII collaborations and a largish and promising region
is completely consistent with those in FIGs.~\ref{CSMfi}~-~\ref{CSMH}. The upcoming XENON1T and SuperCDMS projected limits which are expected to improve by one order of magnitude their limits will be sensitive enough to further restrict our model or reveal its plausibility.

Moreover we have discussed the possibility of explaining $70\%$ up to $100\%$ of 
the Fermi-LAT observed gamma-ray emission from the galactic center through a dominant annihilation into 
$b\bar{b}$ final states with a WIMP mass in the $(25-40)$~GeV range in FIG.~\ref{ThermalCSMfi}, showing
that the 331LHN model has a large amount of parameter space to offer a plausible explanation for these events.

Additionally, we have networked the struggling  probe for WIMPs in underground experiments with the ongoing Higgs boson search at Tevatron and LHC. We have concluded that, in case our WIMP explains the gamma-ray
emission from the galactic center, the Higgs boson decays primarily to a pair of WIMPs. In particular in FIG.~\ref{GammaHfifi}, we have exhibited that, $BR(H \rightarrow 2 WIMPs) \geq 90 \%$, with $BR(H \rightarrow \gamma \gamma) \simeq 2.9 \times 10^{-5} $, $BR(H \rightarrow b \bar{b}) \simeq 1.4 \times 10^{-2} $, and $BR(H \rightarrow \tau \bar{\tau}) \simeq 7.9 \times 10^{-4} $, for the case that $M_{H}=125$~GeV and $M_{WIMP}=20$~GeV.  It is important to emphasize that the branching into a pair of photons in 331LHN, for example, is roughly two orders of magnitude smaller than the SM. Furthermore, we have obtained that in the scenario where $M_{WIMP} < 60$~GeV, we could explain the non-observation of any mild excess at Tevatron and LHC, as well as link the nature of the dark matter with Higgs boson paradigm, which soon will unveil where the physics beyond the SM lies on. This would demand additional efforts on the side of collider search since the branching ratio for Higgs decay into SM particles would be very suppressed in the 331LHN model.
 
Notwithstanding, if the recent modest excesses measured at Tevatron and LHC are confirmed in the near future and they turn out to be consistent with a SM Higgs, this completely rules out the DM hypothesis to explain the gamma-ray emission from GC in our model, once this implies that $ M_{WIMP} > 60 $~GeV and only in this mass range our model recovers the SM predictions.
Lastly, we have shown in FIGs.~\ref{BRHAA}-\ref{BRHtautau} that under the latter hypothesis our model predicts no deviations from the SM Higgs boson decay channels into $\gamma \gamma$, $b \bar{b}$, and $\tau \bar{\tau}$, which are the most promising channels for searching a light Higgs boson.

We also remark that for the regime where $M_{WIMP} < M_Z/2$ no bound can be derived regarding the invisible width of the Z boson, because all decay modes which involve a WIMP in the final state have at least one 331 particle which is much heavier than the Z boson.

\section*{Acknowledgments}

We would like to thank Chris Kelso, Dan Hooper, LianTao, Daniele Alves, A. Semenov as well as A. G. Dias for valuable discussions and/or comments. CASP and PSRS are supported by the Conselho Nacional de Desenvolvimento Cient\'{\i}fico e Tecnol\'ogico (CNPq), FSQ by Coordena\c{c}\~ao de aperfei\c{c}oamento de Pessoal de N\'{\i}vel Superior (CAPES) and  DR has been supported in part by UdeA/2011 grant: IN1614-CE. 

\appendix
\section{}
\label{sec6}

The invisible width $H \rightarrow  WIMP + WIMP$ in our model is given by,

\begin{equation}
\Gamma_{WIMP}= \frac{\lambda_{(H\Phi \Phi)}^2}{32 \pi} \frac{ \sqrt{M_H^2 - 4 M_{WIMP}^2}}{M_H^2},
\label{GammaWIMP}
\end{equation}where,

\begin{equation}
\lambda_{(H\Phi \Phi)}=\frac{-1}{\sqrt{2}(1+\frac{v^2}{V^2})} ( 3\lambda_2 v + \frac{v^3}{2 V^2}+ \lambda_7 \frac{v^3}{V^2} + \lambda_7 v + \frac{v}{2} ).
\label{coupling}
\end{equation}

WIMP refers to the scalar $\Phi$ in the model. Here $v=\frac{v_{SM}}{\sqrt{2}}$, and V is the scale of symmetry breaking of the 331 model, which we assume to be $\gtrsim 1$ TeV. Different values for V produce similar results.

It is important to notice that these couplings in Eq.~(\ref{coupling}) are determined by the mass of the WIMP and Higgs boson, through the following equations,

\begin{equation}
M_{WIMP}^2=\frac{\lambda_7 + 1/2}{2} ( v^2 + V^2 ),
\label{MWIMP}
\end{equation}

\begin{equation}
M_{H}^2=3 \lambda_2 v.
\label{MH}
\end{equation}

Therefore, fixing V in few TeV and plugging Eq.~(\ref{coupling})
~-~\ref{MH} into Eq.~(\ref{GammaWIMP}) we can express the invisible width as a function of the Higgs boson and WIMP masses only. For $M_H \sim 120$~GeV , and $20\ \mbox{GeV} \leq  M_{WIMP} \leq  60$~GeV, the WIMP is the heaviest particle which the Higgs can decay to. For this reason the branching ratio of the Higgs boson into a pair of WIMPs is dominant in this mass range as we may observe in FIG.~(\ref{GammaHfifi}).
{\section*{References}}

\end{document}